\documentclass[conference]{IEEEtran}
\IEEEoverridecommandlockouts

\usepackage{cite}
\usepackage{amsmath,amssymb,amsfonts}
\usepackage{algorithmic}
\usepackage{graphicx}
\usepackage{textcomp}
\usepackage{xcolor}
\usepackage{mathtools,array,makecell, bm}
\def\BibTeX{{\rm B\kern-.05em{\sc i\kern-.025em b}\kern-.08em
    T\kern-.1667em\lower.7ex\hbox{E}\kern-.125emX}}

\begin{document}

\title{Hybrid Losses for Hierarchical Embedding Learning\\
\thanks{Haokun Tian is supported by UK Research and Innovation [grant number EP/S022694/1]. This research utilised Queen Mary's Apocrita HPC facility, supported by QMUL Research-IT. http://doi.org/10.5281/zenodo.438045. Code available at https://github.com/tiianhk/label-metric.}
}

\makeatletter
\newcommand{\linebreakand}{%
  \end{@IEEEauthorhalign}
  \hfill\mbox{}\par
  \mbox{}\hfill\begin{@IEEEauthorhalign}
}
\makeatother

\author{
\IEEEauthorblockN{Haokun Tian$^{1}$ \qquad Stefan Lattner$^{2}$ \qquad Brian McFee$^{3}$ \qquad Charalampos Saitis$^{1}$}
\IEEEauthorblockA{
\textit{$^{1}$Center for Digital Music, Queen Mary University of London, London, UK} \\
\textit{$^{2}$Music Team, Sony Computer Science Laboratories, Paris, France} \\
\textit{$^{3}$Music and Audio Research Laboratory, New York University, New York, USA} \\
haokun.tian@qmul.ac.uk}}


\maketitle

\begin{abstract}
In traditional supervised learning, the cross-entropy loss treats all incorrect predictions equally, ignoring the relevance or proximity of wrong labels to the correct answer. By leveraging a tree hierarchy for fine-grained labels, we investigate hybrid losses, such as generalised triplet and cross-entropy losses, to enforce similarity between labels within a multi-task learning framework. We propose metrics to evaluate the embedding space structure and assess the model's ability to generalise to unseen classes, that is, to infer similar classes for data belonging to unseen categories. Our experiments on OrchideaSOL, a four-level hierarchical instrument sound dataset with nearly 200 detailed categories, demonstrate that the proposed hybrid losses outperform previous works in classification, retrieval, embedding space structure, and generalisation.
\end{abstract}


\renewcommand{\arraystretch}{1.3}

\section{Introduction}\label{sec:introduction}

Deep neural networks trained with human-annotated labels have achieved remarkable performance in classification and retrieval tasks. However, as pointed out in \cite{bertinetto2020making}, despite the vast advances in prediction accuracy since AlexNet \cite{krizhevsky2012imagenet}, models have remained indifferent to false predictions. Put differently, although models are making fewer incorrect predictions with better architectures and larger datasets, the semantics of these predictions have not become closer to the correct answers. Therefore, it is necessary to explicitly incorporate label similarity into the current training and evaluation paradigm. Ideally, this would lead to the model assigning some probability mass to similar classes, as well as having reasonable failure modes. For example, if the model incorrectly predicts the sound of a musical instrument, it should at least predict an instrument with a similar timbre rather than a completely unrelated sound.

Tree hierarchies are a natural way to represent label similarity. They are provided for some large datasets such as ImageNet \cite{deng2009imagenet} and AudioSet \cite{gemmeke2017audio}, as well as in more specialised datasets like UrbanSound \cite{salamon2014dataset}. Established hierarchies can be directly applied to suitable datasets \cite{garcia2021leveraging}. If none exists, creating a tree hierarchy for a set of labels using domain knowledge can be relatively efficient. However, mainstream representation learning research commonly ignores label similarities, training and evaluating models solely on fine-grained categories where wrong labels are treated equally \cite{kong2020panns, gong21b_interspeech, chen2022hts}. To incorporate label hierarchies, Zhang et al. \cite{zhang2016embedding} proposed a multi-task learning framework that integrates a generalised triplet loss—designed to learn an embedding space reflecting the hierarchical structure of the data—with a softmax loss for classification, which maps the embedding to produce the probabilities of each fine-grained category. This work has been applied to audio event detection by Jati et al. \cite{jati2019hierarchy}. In addition to \cite{zhang2016embedding}, \cite{jati2019hierarchy} also trained a classifier for the superclasses, i.e., the parents of the fine-grained classes, but this was only for comparison and was not used in the multi-task learning. Other approaches focus solely on classification \cite{bertinetto2020making, krause2022hierarchical, krause2023hierarchical} or embedding learning \cite{nolasco2022rank}. Among them, \cite{krause2022hierarchical, krause2023hierarchical} jointly predicted classes across different hierarchy levels using the binary cross-entropy loss. In this work, we extend the more versatile multi-task learning framework introduced by Zhang et al. with the following contributions:

\begin{itemize}
    \item We introduce hybrid hierarchy-aware classification losses that integrate the prediction of all valid parent nodes of the respective data point into the multi-task framework. This approach achieves better organisation of the embedding space compared to prior work \cite{zhang2016embedding} and \cite{jati2019hierarchy}, which only consider classification on the fine-grained level.
    \item A set of metrics is introduced to evaluate how well the embedding space structure matches the tree hierarchy and evaluates the model's ability to generalise to unseen classes. For the latter, we estimate whether the model's output is similar to the true class. This is in contrast to \cite{zhang2016embedding, jati2019hierarchy, nolasco2022rank} where evaluations are performed per tree level (e.g., a coarse level and a fine level for a bi-level hierarchy), and no unseen classes are used for testing.
\end{itemize}

Our approach can serve as a blueprint for other tasks that benefit from hierarchical embeddings, including few-shot audio event recognition \cite{garcia2021leveraging}, audio source separation \cite{petermann2023hyperbolic}, and music structure analysis \cite{buisson2024self}. Also, with the model's inter-label generalisation ability improved, it can better handle in-the-wild data by linking new examples to familiar (and similar) labels. 

Apart from traditional supervised learning, we recognise that using natural language supervision to learn representations for audio \cite{elizalde2023clap, wu2023large} and image \cite{radford2021learning} is another scheme to capture label similarity. While these models demonstrate strong zero-shot ability, they still lag behind traditional supervised learning on some downstream classification tasks \cite{elizalde2023clap}. Thus, equipping traditional supervised learning models with the ability to capture label similarity remains valuable.

\section{Methods}\label{sec:method}

We consider a dataset $X=\{x_i\}_{i=1}^N$ with the label set $Y$ and construct a tree hierarchy $T$ for labels. Each data sample $x_i$ is initially associated with a label from $Y$, and each label starts as a leaf node. The leaf nodes are grouped based on label similarity, with each group connected to a parent node. Some nodes may stay unconnected and join the next parent nodes for further grouping. This process is repeated iteratively until all nodes are connected up to a single ancestor node, also known as the root node. As a parent node concludes the labels of its children, we define a mapping $\mathcal{M}: \mathbf{n} \rightarrow X_\mathbf{n}$, where $\mathbf{n}$ is a node in $T$ and $X_\mathbf{n}$ is a subset of $X$ containing all data examples associated with the labels of the child nodes under $\mathbf{n}$. From here, we use ``sample from node $\mathbf{n}$" to denote ``sample from $\mathcal{M}(\mathbf{n})$".

\subsection{Triplet Learning}

Triplet learning is a sub-field in metric learning, which aims to learn an embedding space where similar items are close to each other and dissimilar items are pushed away from each other \cite{schroff2015facenet}. Triplet loss (section \ref{sec:loss}), operating on data triplet $(x_a, x_p, x_n)$ sampled from the dataset $X$, requires the positive $x_p$ to be more similar to the anchor $x_a$ than the negative $x_n$. Sampling (mining) finds the most informative triplets to compute the training loss, sidestepping easy triplets that contribute zero loss values \cite{wu2017sampling}. Offline mining \cite{harwood2017smart} prepares triplets before training batch construction and online mining constructs the triplets from a given batch \cite{schroff2015facenet, wu2017sampling}.

\subsection{Sampling Triplets from the Tree}\label{sec:sample}

Generalised triplets \cite{zhang2016embedding} are triplets with different similarity levels. An example is that a string instrument is dissimilar to a brass instrument on the \emph{instrument family level}, and within the string family, a violin is dissimilar to a viola on the \emph{instrument level}. This extends the standard way of choosing triplets, where $x_a$ and $x_p$ are from the same leaf node and $x_n$ is from another. With the tree hierarchy, one can sample $x_a$ and $x_p$ from different leaf nodes and sample $x_n$ from outside their LCA (lowest common ancestor). For example, it is valid to sample a violin as the anchor, a viola as the positive, and a brass instrument as the negative.

To describe our sampling process for any tree structure, we use $\mathbf{n}_a$, $\mathbf{n}_p$, and $\mathbf{n}_n$ to represent the triplet, and $\mathbf{n}_+$ to denote the parent node of $\mathbf{n}_a$ and $\mathbf{n}_p$. Iterating over parent nodes with more than one child in $T$, we choose all possible children pairs under the parent to form ($\mathbf{n}_+$, $\mathbf{n}_n$). If $\mathbf{n}_+$ only has one child, we randomly sample the anchor and the positive from it. If $\mathbf{n}_+$ has more than one child, we again choose all possible children pairs under it to obtain ($\mathbf{n}_a$, $\mathbf{n}_p$). This process constructs the training data for an epoch and can be seen as an offline triplet mining technique implemented as a PyTorch Sampler.

\subsection{Hierarchy-Aware Losses}\label{sec:loss}

We first review the triplet loss and the cross-entropy losses. The triplet loss requires a margin $\alpha$ between the distance of $x_a$ and $x_p$ and the distance of $x_a$ and $x_n$ in the embedding space. We use the negative cosine similarity to compute the distance between two embeddings:
\begin{equation}\label{eq:triplet}
\mathcal{L}_{\text{triplet}}(x_a,x_p,x_n)=\max (0,d_{ap}-d_{an}+\alpha).
\end{equation}

The cross-entropy losses are computed between the predicted categorical probability distributions and the true labels. The raw model outputs are mapped to probabilities $\hat{y}$ using the sigmoid function for binary classification and the softmax function for multi-class classification. The loss is defined between the prediction $\hat{y}$ and the true label $y$:
\begin{equation}\label{eq:binary}
\mathcal{L}_{\text{binary}}(\hat{y},y) = - [ y \cdot \log(\hat{y}) + (1 - y) \cdot \log(1 - \hat{y}) ],
\end{equation}
\begin{equation}\label{eq:multiclass}
\mathcal{L}_{\text{multi-class}}(\hat{y},y) = - \sum_{i=1}^{C} y_i \cdot \log(\hat{y}_i),
\end{equation}
where $y$ is 0 (false) or 1 (true) for binary classification, one-hot encoded vectors for multi-class classification, and $C$ is the number of classes. Below, we introduce how \eqref{eq:triplet}-\eqref{eq:multiclass} are employed to capture tree-structured label spaces. All losses simplify to their standard versions when the tree hierarchy becomes flat.

\textbf{Triplet Loss} (\texttt{T}): the loss \eqref{eq:triplet} is computed on triplets obtained from the hierarchical sampling process that retrieves positive pairs from different  levels of the hierarchy (section \ref{sec:sample}). Different from \cite{zhang2016embedding, jati2019hierarchy} where different margins for different similarity levels are used, we advocate using a single margin so that the number of margins does not scale with the depth of the tree. However, the dimensionality of the embeddings should scale with increasing tree size, guaranteeing a large enough representation space.

\begin{figure}[t]
  \centering
  \includegraphics[width=0.48\textwidth]{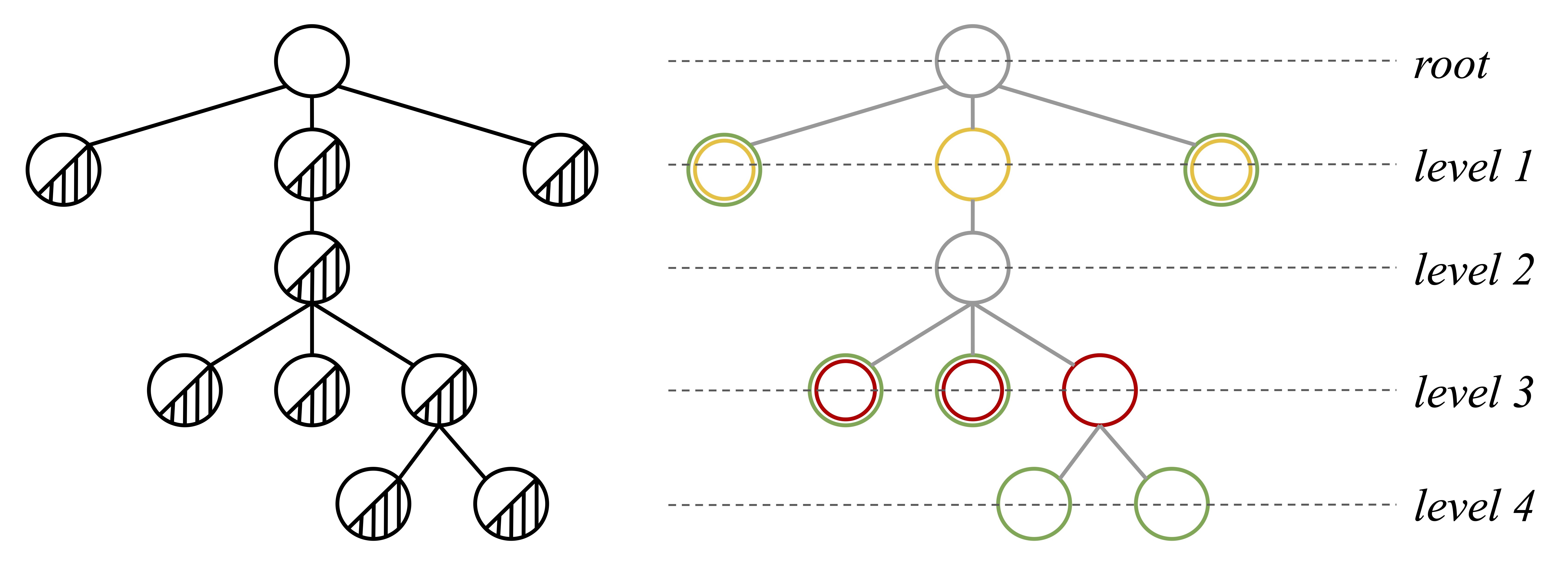}
  \caption{Visualisation of how targets are chosen from a sample tree for different hierarchy-aware losses. Left: shading indicates binary classification on the node. Right: multi-class classification is performed within each group of nodes sharing the same colour.}
  \label{fig:loss}
\end{figure}

\textbf{Binary Loss} (\texttt{B}): similar to \cite{krause2022hierarchical, krause2023hierarchical}, standard binary classification is performed on each node in the tree except the root node (see Fig. \ref{fig:loss} left). This constitutes a multi-label classification (tagging) task where each sample is labeled by a leaf node and its corresponding ancestors along the path to the root. The loss for an arbitrary sample $x \in X$ is computed as:
\begin{equation}
    \frac{1}{|T|-1}\sum\limits_{j=1}^{|T|-1}\mathcal{L}_{\text{binary}}(\hat{y}_{\mathbf{n}_j},y_{\mathbf{n}_j}),
\end{equation}
where $|T|$ is the number of nodes in $T$, $\hat{y}_{\mathbf{n}_j}$ is the predicted probability for node $\mathbf{n}_j$, and $y_{\mathbf{n}_j}$ denotes whether $x \in \mathcal{M}(\mathbf{n}_j)$.

\textbf{Per-Level Loss} (\texttt{PL}): standard multi-class classification is performed on each level of the tree. Taking the tree structure in Fig. \ref{fig:loss} as an example, levels with only one node are omitted (root and level 2). Nodes on levels 1 and 3 are marked in yellow and red respectively. Nodes on level 4 are joined by other leaf nodes and are marked in green, corresponding to all the fine-grained labels in $Y$. We additionally denote this loss for all leaf nodes as \texttt{L} (\textbf{Leaf Loss}). As there are three levels with more than one node, a total of three classification tasks are created from the tree in Fig. \ref{fig:loss}. The loss is computed as:
\begin{equation}
\sum\limits_{j=1}^{N}\mathcal{L}_{\text{multi-class}}(\hat{y}_{\ell_j},y_{\ell_j}),
\end{equation}
where $N$ is the number of target levels, $\hat{y}_{\ell_j}$ is the predicted probability distribution on the $j$-th level, and $y_{\ell_j}$ is the corresponding one-hot target.

\section{Experiments and Evaluation}\label{sec:experiment}

\subsection{Dataset}

We use the OrchideaSOL dataset \cite{cella2020orchideasol}. It contains recordings of single musical instrument notes with a four-level hierarchy. The four levels indicate the instrument family, the instrument, the type of mute, and the playing technique. Mutes are accessories commonly used in brass and string instruments to alter the timbre or soften the volume. For example, to mute a trumpet, a device is inserted into its bell. Playing techniques further create timbre variations with different player interactions. For example, when playing `pizzicato', a player plucks the strings of an instrument that is typically bowed. The four levels contain 5, 16, 35, and 197 categories, respectively, totaling a dataset size of $\sim$13K instances.

\subsection{Data Split}

We split leaf nodes into ``seen nodes'' and ``unseen nodes'' (i.e., seen or unseen during training), where leaf nodes with less than 10 samples are directly considered ``unseen'' due to data shortage. The remaining leaves are partitioned using 5-fold cross-validation, making 20\% of them unseen during training. For each seen leaf node, we split data within the node (i.e., split $\mathcal{M}(\mathbf{n})$ for $\mathbf{n}$) into train, valid, and test set by 8:1:1. This yields $\sim$8K training samples, $\sim$1K validation samples, $\sim$1K test samples, and $\sim$3K prediction samples (whose classes are unseen).

\begin{figure}[t]
  \centering
  \includegraphics[width=0.48\textwidth]{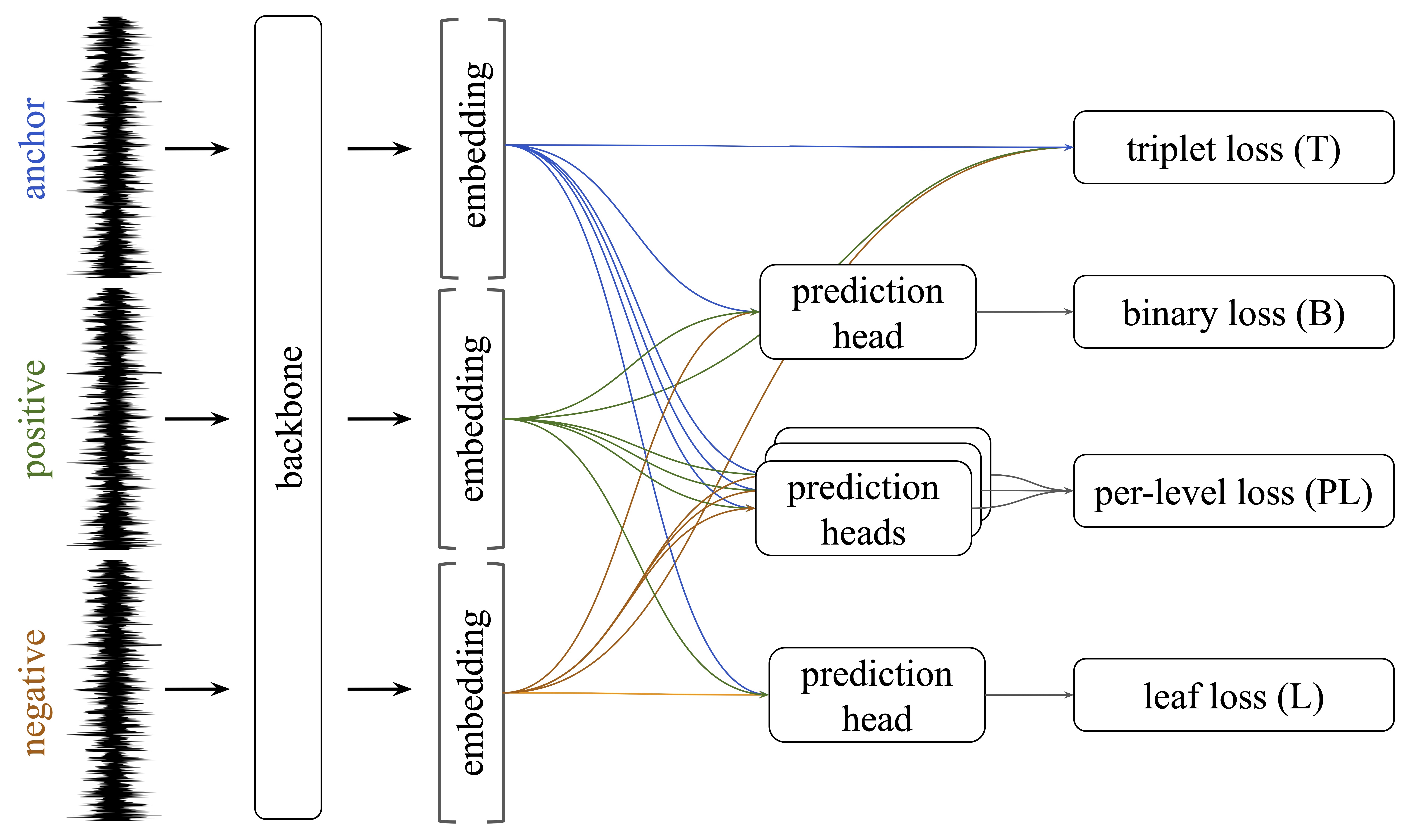}
  \caption{Overview of the model pipeline.}
  \label{fig:model}
\end{figure}

\subsection{Training}

Model input is one-second audio with a sampling rate of 44.1kHz. Audio samples shorter than one second are zero-padded, and those longer are truncated (to the right). Around 2.2K triplets are sampled at the start of each training epoch (the number varies across folds as it is affected by the tree structure). We compute the dB-scaled Mel spectrogram with $\texttt{n\_fft}=2048$ ($\sim$46ms), $\texttt{hop\_length}=512$ ($\sim$12ms), and $\texttt{n\_mels}=128$. We use the same backbone CNN architecture as \cite{garcia2021leveraging}, with an (output) embedding size of 256. These embeddings are passed to the linear prediction heads to produce raw, unnormalised probability scores for the losses (Fig. \ref{fig:model}). We test six loss combinations for multi-task learning, namely \texttt{L}, \texttt{L+T}, \texttt{PL}, \texttt{PL+T}, \texttt{PL+B}, and \texttt{PL+B+T}. For the triplet loss, we set the margin $\alpha=0.3$. For classification losses (\texttt{L}, \texttt{B} and \texttt{PL}), class weights are set inversely proportional to their sampling frequency. Losses are combined using uniform weighting, which is compared against other weighting strategies in \cite{gong2019comparison}. We use a batch size of 32.

\subsection{Evaluation}\label{sec:evaluation}

\begin{table*}[htbp]
\caption{Results evaluated on the Test and  Prediction sets.}
\centering
\begin{tabular}{c|cccc|cccc}
\Xhline{2\arrayrulewidth}
& \multicolumn{4}{c|}{\textbf{Test Set}} & \multicolumn{4}{c}{\textbf{Prediction Set}} \\ \hline

& \textbf{\textit{leaf F1}} & \textbf{\textit{leaf RP@5}} & \textbf{\textit{MNR}} $\downarrow^{\mathrm{*}}$ & \textbf{\textit{NDCG}} & $\textbf{\textit{Acc}}_{\textbf{\textit{blind}}}$ & $\textbf{\textit{Acc}}_{\textbf{\textit{aware}}}$ & $\textbf{\textit{Acc}}_{\textbf{\textit{blind}}} / \textbf{\textit{Acc}}_{\textbf{\textit{aware}}}$ & \textbf{\textit{NDCG}} \\ \hline

\texttt{L} & $96.4\,(0.2)$ & $60.4\,(0.4)$ & $19.5\,(0.2)$ & $92.4\,(0.1)$ & $58.2\,(2.5)$ & $-$ & $-$ & $95.8\,(0.4)$ \\ 

\texttt{L+T} & $96.6\,(0.2)$ & $62.3\,(3.1)$ & $17.3\,(0.7)$ & $93.5\,(0.5)$ & $60.2\,(2.5)$ & $-$ & $-$ & $96.1\,(0.6)$ \\ 

\texttt{PL} & $96.9\,(0.2)$ & $62.1\,(0.3)$ & $\mathbf{11.2\,(0.3)}$ & $\mathbf{96.6\,(0.2)}$ & $62.6\,(2.7)$ & $79.4\,(2.7)$ & $78.9\,(1.6)$ & $\mathbf{97.7\,(0.2)}$ \\ 

\texttt{PL+T} & $96.7\,(0.3)$ & $67.1\,(2.4)$ & $11.7\,(0.2)$ & $96.2\,(0.2)$ & $65.5\,(2.3)$ & $79.2\,(2.9)$ & $82.7\,(1.4)$ & $97.2\,(0.2)$ \\ 

\texttt{PL+B} & $96.9\,(0.3)$ & $60.2\,(1.0)$ & $11.9\,(0.3)$ & $95.6\,(0.2)$ & $63.5\,(2.5)$ & $\mathbf{81.1\,(3.4)}$ & $78.4\,(1.7)$ & $97.4\,(0.3)$ \\ 

\texttt{PL+B+T} & $\mathbf{97.2\,(0.2)}$ & $\mathbf{67.2\,(2.2)}$ & $13.1\,(0.3)$ & $95.3\,(0.3)$ & $\mathbf{65.7\,(2.3)}$ & $79.1\,(2.6)$ & $\mathbf{83.2\,(1.6)}$ & $96.7\,(0.3)$ \\ \Xhline{2\arrayrulewidth}

\multicolumn{9}{l}{All numbers are in percentages. $^{\mathrm{*}}$ indicates that a lower value is better. \texttt{L}, \texttt{T}, \texttt{PL}, and \texttt{B} denote leaf, triplet, per-level, and binary loss.}
\end{tabular}
\label{table}
\end{table*}

\subsubsection{Classification}\label{sec:eval_classification}

We report the standard F1-score for the leaf node prediction heads of \texttt{L} and \texttt{PL} on the test set (leaf F1). In addition, to determine whether the model makes reasonable errors on unseen classes (i.e., predicting instruments with timbres similar to the unseen classes), we introduce the concept of the Lowest Seen Ancestor (LSA). The LSA is defined as the first ancestor of an unseen leaf node that belongs to a class the model has encountered, tracing the path from the leaf to the root. We report two accuracies $\text{Acc}_{\text{blind}}$ and $\text{Acc}_{\text{aware}}$ for predicting the LSA of their corresponding unseen leaves of data samples in the prediction set. For the \textit{blind} case, we start from the model's incorrect leaf prediction and trace up to the level of the true LSA to obtain a prediction, which reflects whether the incorrect prediction is a descendant of the true yet unseen leaf's LSA. For the \textit{aware} case, we use the prediction head in \texttt{PL} that is on the same level as the true LSA for prediction, which makes LSA a seen class during training and serves as a reference for the \textit{blind} case as they predict from the same embeddings. The ratio between two accuracies is presented where applicable.

\subsubsection{Retrieval}

We report the standard \textbf{R}etrieval \textbf{P}recision @ 5 for leaf nodes (leaf RP@5) and three ranking-based metrics (MNR, $\text{NDCG}_{\text{sum}}$, $\text{NDCG}_{\text{max}}$) that assess how well the embedding space aligns with a pre-defined tree hierarchy. Given a query $q_i$, candidates are ranked in descending order based on cosine similarity in the embedding space.

\textbf{MNR}: we modify the \textbf{M}ean \textbf{N}ormalised \textbf{R}ank \cite{lattner2022samplematch, torres2023singer} to take into account ranks of similar samples on different tree levels. On the path from $q_i$'s leaf node to the root, samples from the $j$-th node $\mathbf{n}_{ij}$ are considered similar to the query. Levels with only one node are skipped in the same way as \texttt{PL}. The metric is computed as:
$$\text{MNR}=\frac{1}{Q}\sum\limits_{i=1}^Q\left(\frac{1}{L}\sum\limits_{j=1}^{L}\left(\frac{1}{M_{ij}}\sum\limits_{k=1}^{M_{ij}}\frac{r_{ijk} - 1}{N}\right)\right),$$
where $Q$ is the number of queries, $L$ is the number of tree levels with more than one node (invariant to the query), $M_{ij}$ is the number of correct answers, and $r_{ijk}$ is the rank of the correct answer, shifted by 1 and normalised by $N$ (the number of candidates) to be in the range $[0,1)$. We highlight that more weights are applied naturally to correct answers on deeper levels (further from the root), as the number of times a candidate is counted as correct is the number of nodes it belongs to in the list of $q_i$'s leaf node and its ancestors.

\textbf{NDCG}: the \textbf{N}ormalised \textbf{D}iscounted \textbf{C}umulative \textbf{G}ain is a standard metric that evaluates the quality of predicted rankings by comparing it to ideal rankings. The true relevance between samples, which determines the ideal ranking, is derived from the distance on the tree between their corresponding leaf nodes. With the tree height denoted as $H_T$ (number of edges on the longest path from the root to a leaf node) and the tree diameter as $D_T$ (number of edges on the longest path between two leaf nodes), we define two node-wise relevance scores considering two types of node distance:
$$rel_{\text{sum}}(\mathbf{n}_1, \mathbf{n}_2)=1-\frac{d(\mathbf{n}_1, \mathbf{n}_{\text{LCA}}) + d(\mathbf{n}_2, \mathbf{n}_{\text{LCA}})}{D_T},$$
$$rel_{\text{max}}(\mathbf{n}_1, \mathbf{n}_2)=1-\frac{\max (d(\mathbf{n}_1, \mathbf{n}_{\text{LCA}}), d(\mathbf{n}_2, \mathbf{n}_{\text{LCA}}))}{H_T},$$
where $\mathbf{n}_{\text{LCA}}$ refers to the lowest common ancestor of $\mathbf{n}_1$ and $\mathbf{n}_2$, and $d(\cdot)$ counts the number of edges on the path from the descendant to the ancestor. $\text{NDCG}_{\text{sum}}$ and $\text{NDCG}_{\text{max}}$ are computed accordingly.

\section{Results and Discussion}\label{sec:results}

Table \ref{table} presents the mean and the standard error of the mean for evaluation metrics computed on five test and prediction folds. In all experiments, $\text{NDCG}_{\text{sum}}$ and $\text{NDCG}_{\text{max}}$ are identical up to the fourth decimal place therefore we report only one $\text{NDCG}$ value.

Previous work \cite{zhang2016embedding, jati2019hierarchy} showed that \texttt{L+T} is superior to only \texttt{L} for classification and retrieval tasks. The results in this section show that our proposed hybrid losses (\texttt{PL}, \texttt{PL+T}, \texttt{PL+B}, \texttt{PL+B+T}) further improve performance. Among all metrics, MNR and NDCG indicate how well the embedding space \emph{globally} reflects the hierarchical taxonomies and are, therefore, strongly correlated. Here, \texttt{PL} alone delivers the best results on the test and prediction sets. This is reasonable, as the triplet loss \texttt{T} operates on a rather local scale (through the threshold parameter) and focuses more on finer levels due to the tree-structure-sensitive sampling. The binary loss \texttt{B} captures less hierarchical information, as it handles labels individually and no inter-label comparison is involved. Overall, losses including \texttt{PL} outperform the baselines \texttt{L} and \texttt{L+T} on the test set by around $4\sim8\%$ in MNR and $2\sim4\%$ in NDCG. We observe higher absolute values of NDCG on the prediction set. This is due to the number of leaf nodes in the prediction set being only one-fourth of the number of leaf nodes in the test set, forming a more sparse tree that is easier to match.

For the fine-grained classification metric leaf F1, new hybrid losses show improvements over \texttt{L} and \texttt{L+T}. This suggests that learning the hierarchy as additional information helps the model predict more accurately, which is in line with the aim of other representation learning methods. For the fine-grained retrieval metric leaf RP@5, \texttt{PL+T} and \texttt{PL+B+T} outperform other losses by $5\%$, with \texttt{PL} capturing the global structure and \texttt{T} aggregating samples of each leaf node. Adding \texttt{B} brings only a $0.1\%$ increase, which is minimal and suggests that most useful structures are already captured by \texttt{PL} and \texttt{T}. On the prediction set, the accuracy of predicting the LSA (lowest seen ancestor) reflects the model's ability to generalise to unseen classes. Again, \texttt{PL+T} and \texttt{PL+B+T} make the most relevant leaf guesses compared to other losses. Their prediction ($\text{Acc}_{\text{blind}}$) achieves over $80\%$ of the performance of supervised classification ($\text{Acc}_{\text{aware}}$) using the same embeddings.

Expanding on the multi-task learning framework with hybrid losses, our future work includes identifying the optimal triplet sampling strategy for tree structures, investigating if online mining further improves triplet selection, adapting the current approach to hyperbolic embeddings \cite{nickel2017poincare, khrulkov2020hyperbolic}, and applying our methods to large multi-label datasets, such as AudioSet, for general-purpose model pre-training.

\vfill\pagebreak

\bibliographystyle{IEEEtran}
\bibliography{refs}

\begin{thebibliography}{10}
\providecommand{\url}[1]{#1}
\csname url@samestyle\endcsname
\providecommand{\newblock}{\relax}
\providecommand{\bibinfo}[2]{#2}
\providecommand{\BIBentrySTDinterwordspacing}{\spaceskip=0pt\relax}
\providecommand{\BIBentryALTinterwordstretchfactor}{4}
\providecommand{\BIBentryALTinterwordspacing}{\spaceskip=\fontdimen2\font plus
\BIBentryALTinterwordstretchfactor\fontdimen3\font minus \fontdimen4\font\relax}
\providecommand{\BIBforeignlanguage}[2]{{%
\expandafter\ifx\csname l@#1\endcsname\relax
\typeout{** WARNING: IEEEtran.bst: No hyphenation pattern has been}%
\typeout{** loaded for the language `#1'. Using the pattern for}%
\typeout{** the default language instead.}%
\else
\language=\csname l@#1\endcsname
\fi
#2}}
\providecommand{\BIBdecl}{\relax}
\BIBdecl

\bibitem{bertinetto2020making}
L.~Bertinetto, R.~Mueller, K.~Tertikas, S.~Samangooei, and N.~A. Lord, ``Making better mistakes: Leveraging class hierarchies with deep networks,'' in \emph{CVPR}, 2020.

\bibitem{krizhevsky2012imagenet}
A.~Krizhevsky, I.~Sutskever, and G.~E. Hinton, ``Imagenet classification with deep convolutional neural networks,'' in \emph{NeurIPS}, 2012.

\bibitem{deng2009imagenet}
J.~Deng, W.~Dong, R.~Socher, L.-J. Li, K.~Li, and L.~Fei-Fei, ``Imagenet: A large-scale hierarchical image database,'' in \emph{CVPR}, 2009.

\bibitem{gemmeke2017audio}
J.~F. Gemmeke, D.~P. Ellis, D.~Freedman, A.~Jansen, W.~Lawrence, R.~C. Moore, M.~Plakal, and M.~Ritter, ``Audio set: An ontology and human-labeled dataset for audio events,'' in \emph{ICASSP}, 2017.

\bibitem{salamon2014dataset}
J.~Salamon, C.~Jacoby, and J.~P. Bello, ``A dataset and taxonomy for urban sound research,'' in \emph{ACM MM}, 2014.

\bibitem{garcia2021leveraging}
H.~F. Garcia, A.~Aguilar, E.~Manilow, and B.~Pardo, ``Leveraging hierarchical structures for few-shot musical instrument recognition,'' in \emph{ISMIR}, 2021.

\bibitem{kong2020panns}
Q.~Kong, Y.~Cao, T.~Iqbal, Y.~Wang, W.~Wang, and M.~D. Plumbley, ``Panns: Large-scale pretrained audio neural networks for audio pattern recognition,'' \emph{IEEE ACM Trans. Audio Speech Lang. Process.}, 2020.

\bibitem{gong21b_interspeech}
Y.~Gong, Y.-A. Chung, and J.~Glass, ``{AST: Audio Spectrogram Transformer},'' in \emph{Interspeech}, 2021.

\bibitem{chen2022hts}
K.~Chen, X.~Du, B.~Zhu, Z.~Ma, T.~Berg-Kirkpatrick, and S.~Dubnov, ``Hts-at: A hierarchical token-semantic audio transformer for sound classification and detection,'' in \emph{ICASSP}, 2022.

\bibitem{zhang2016embedding}
X.~Zhang, F.~Zhou, Y.~Lin, and S.~Zhang, ``Embedding label structures for fine-grained feature representation,'' in \emph{CVPR}, 2016.

\bibitem{jati2019hierarchy}
A.~Jati, N.~Kumar, R.~Chen, and P.~Georgiou, ``Hierarchy-aware loss function on a tree structured label space for audio event detection,'' in \emph{ICASSP}, 2019.

\bibitem{krause2022hierarchical}
M.~Krause and M.~M{\"u}ller, ``Hierarchical classification of singing activity, gender, and type in complex music recordings,'' in \emph{ICASSP}, 2022.

\bibitem{krause2023hierarchical}
------, ``Hierarchical classification for instrument activity detection in orchestral music recordings,'' \emph{IEEE ACM Trans. Audio Speech Lang. Process.}, 2023.

\bibitem{nolasco2022rank}
I.~Nolasco and D.~Stowell, ``Rank-based loss for learning hierarchical representations,'' in \emph{ICASSP}, 2022.

\bibitem{petermann2023hyperbolic}
D.~Petermann, G.~Wichern, A.~Subramanian, and J.~Le~Roux, ``Hyperbolic audio source separation,'' in \emph{ICASSP}, 2023.

\bibitem{buisson2024self}
M.~Buisson, B.~McFee, S.~Essid, and H.~C. Crayencour, ``Self-supervised learning of multi-level audio representations for music segmentation,'' \emph{IEEE ACM Trans. Audio Speech Lang. Process.}, 2024.

\bibitem{elizalde2023clap}
B.~Elizalde, S.~Deshmukh, M.~Al~Ismail, and H.~Wang, ``Clap learning audio concepts from natural language supervision,'' in \emph{ICASSP}, 2023.

\bibitem{wu2023large}
Y.~Wu, K.~Chen, T.~Zhang, Y.~Hui, T.~Berg-Kirkpatrick, and S.~Dubnov, ``Large-scale contrastive language-audio pretraining with feature fusion and keyword-to-caption augmentation,'' in \emph{ICASSP}, 2023.

\bibitem{radford2021learning}
A.~Radford, J.~W. Kim, C.~Hallacy, A.~Ramesh, G.~Goh, S.~Agarwal, G.~Sastry, A.~Askell, P.~Mishkin, J.~Clark \emph{et~al.}, ``Learning transferable visual models from natural language supervision,'' in \emph{ICML}, 2021.

\bibitem{schroff2015facenet}
F.~Schroff, D.~Kalenichenko, and J.~Philbin, ``Facenet: A unified embedding for face recognition and clustering,'' in \emph{CVPR}, 2015.

\bibitem{wu2017sampling}
C.-Y. Wu, R.~Manmatha, A.~J. Smola, and P.~Krahenbuhl, ``Sampling matters in deep embedding learning,'' in \emph{ICCV}, 2017.

\bibitem{harwood2017smart}
B.~Harwood, V.~Kumar~BG, G.~Carneiro, I.~Reid, and T.~Drummond, ``Smart mining for deep metric learning,'' in \emph{ICCV}, 2017.

\bibitem{cella2020orchideasol}
C.~E. Cella, D.~Ghisi, V.~Lostanlen, F.~L{\'e}vy, J.~Fineberg, and Y.~Maresz, ``Orchideasol: a dataset of extended instrumental techniques for computer-aided orchestration,'' in \emph{ICMC}, 2020.

\bibitem{gong2019comparison}
T.~Gong, T.~Lee, C.~Stephenson, V.~Renduchintala, S.~Padhy, A.~Ndirango, G.~Keskin, and O.~H. Elibol, ``A comparison of loss weighting strategies for multi task learning in deep neural networks,'' \emph{IEEE Access}, 2019.

\bibitem{lattner2022samplematch}
S.~Lattner, ``Samplematch: Drum sample retrieval by musical context,'' in \emph{ISMIR}, 2022.

\bibitem{torres2023singer}
B.~Torres, S.~Lattner, and G.~Richard, ``Singer identity representation learning using self-supervised techniques,'' in \emph{ISMIR}, 2023.

\bibitem{nickel2017poincare}
M.~Nickel and D.~Kiela, ``Poincar{\'e} embeddings for learning hierarchical representations,'' in \emph{NeurIPS}, 2017.

\bibitem{khrulkov2020hyperbolic}
V.~Khrulkov, L.~Mirvakhabova, E.~Ustinova, I.~Oseledets, and V.~Lempitsky, ``Hyperbolic image embeddings,'' in \emph{CVPR}, 2020.

\end{thebibliography}

\end{document}